\documentclass[pdftex,twocolumn,epjc3]{svjour3}       

\RequirePackage[T1]{fontenc}

\smartqed  

\RequirePackage{graphicx}
\RequirePackage{mathptmx}      
\RequirePackage{flushend}
\RequirePackage[numbers,sort&compress]{natbib}
\RequirePackage[colorlinks,citecolor=blue,urlcolor=blue,linkcolor=blue]{hyperref}

\usepackage[modulo]{lineno}
\journalname{Eur. Phys. J. C}

\begin{document}


\title{Light-flavor squark reconstruction at CLIC$^{*}$}

\author{Frank Simon\thanksref{e1,addr1}
        \and
        Lars Weuste\thanksref{addr1}
}

\thankstext{e1}{e-mail: fsimon@mpp.mpg.de}

\institute{Max-Planck-Institut f\"ur Physik, F\"ohringer Ring 6, 80805 Munich, Germany \label{addr1}}

\date{\today}

\maketitle

\begin{abstract}
We present a simulation study of the prospects for the mass measurement of TeV-scale
light-flavored right-handed squarks at a 3 TeV $e^+e^-$ collider based on CLIC technology. In the considered model, these particles decay into their standard-model counterparts and the lightest neutralino, resulting in a signature of two jets plus missing energy. The analysis is based on
full GEANT4 simulations of the CLIC\_ILD detector concept, including Standard
Model physics backgrounds and beam-induced  ha\-dronic backgrounds from two-photon
processes. The analysis serves as a generic benchmark for the reconstruction of highly energetic jets in events with substantial missing energy. Several jet finding algorithms were evaluated, with the longitudinally invariant $k_t$ algorithm showing a high degree of robustness towards beam-induced background while preserving the features typically found in algorithms developed for $e^+e^-$ collisions. The presented study of the reconstruction of light-flavored squarks shows that for TeV-scale squark masses, sub-percent accuracy on the mass measurement can be a\-chieved at CLIC. 
\end{abstract}

\section{Introduction}
\label{sec:Intro}

{\let\thefootnote\relax\footnote{$^{*}$This work was carried out in the framework of the CLICdp Collaboration}}

Future high energy $e^+e^-$ colliders are precision tools for the discovery and the spectroscopy of new particles expected beyond the Standard Model. One attractive extension of the Standard Model is Supersymmetry \cite{Wess:1973kz, Wess:1974tw}, which predicts a rich spectrum of new particles, one superpartner for each Standard Model particle. These new particles are expected to have masses in the range from about 100 GeV to a few TeV, and may thus come within reach of modern colliders. In the first run of the LHC, no evidence for supersymmetric particles was observed, putting substantial constraints on the allowed parameter space. In particular the mass limits for strongly interacting particles are already high, reaching beyond 1 TeV in certain scenarios such as the cMSSM \cite{Aad:2014wea}. For $R$-parity conserving models, the decay of such particles results in the production of one or more Standard Model particles plus the lightest supersymmetric particle, which typically is a neutralino that escapes undetected \cite{Martin:1997ns}. In analyses based on simplified models \cite{Alwall:2008ag, Alves:2011wf}, mass-degenerate first- and second-generation squarks decaying directly into one quark and the lightest neutralino are excluded up to masses of \mbox{850 GeV} by ATLAS \cite{Aad:2014wea} and up to masses of 875 GeV by CMS \cite{Khachatryan:2015vra}. If such particles exist above those bounds, accessing them in $e^+e^-$ annihilation requires a multi-TeV collider. 

In this article, we study the capabilities of a linear $e^+e^-$ collider based on Compact Linear Collider (CLIC) technology \cite{CLICCDR_vol1} to measure the mass of first- and second-generation right-handed squarks with TeV-scale masses.  In the supersymmetric (SUSY) model used here, these squarks decay almost exclusively through two-body decays into their standard-model counterpart and the lightest neutralino. While the studies are performed within a concrete SUSY scenario used for the physics benchmark studies in the framework of the CLIC conceptual design report (CDR) \cite{Linssen:2012hp}, they can be taken as a more general indication of the CLIC capabilities for pair-produced high-mass states of the same quantum numbers decaying hadronically into a pair of jets and invisible heavy particles. 

\section{Experimental conditions and detectors at CLIC}
\label{sec:ExperimentalConditions}

CLIC is a collider concept based on normal conducting accelerating cavities and two-beam acceleration \cite{CLICCDR_vol1}. It is designed to provide collision energies up to 3 TeV with high luminosity. The project is foreseen to be implemented in several stages \cite{Lebrun:2012hj} to provide optimal luminosity conditions over the whole range of energies, with the final stage reaching the full design energy of the collider. 

At 3 TeV, the experimental conditions are characterized by a luminosity spectrum with a sharp peak at the nominal collision energy and a long tail towards lower energies due to beamstrahlung and initial state radiation, with 35\% of the total luminosity in the top 1\% of the collision energy. Still, most of the available luminosity is concentrated at high energy, so that for processes with a threshold below the maximum energy of the collider the main consequence of the luminosity spectrum is a smearing of the production energy, reducing the effectiveness of energy constraints in the data analysis and resulting in distortions of final state particle energy distributions. 

The interaction of real and virtual photons from the colliding bunches results in particle production through two-photon processes. Of particular relevance for the present analysis is the production of hadrons, $\gamma\gamma \rightarrow$ hadrons, which pre\-sents an important background at CLIC. At 3 TeV, 3.2 events with a $\gamma\gamma$ center-of-mass energy in excess of 2 GeV are expected per bunch crossing within the detector acceptance. Over a full bunch train of 312 bunch crossings, corresponding to 156 ns, a total of approximately 20 TeV are deposited in the calorimeters of a detector at CLIC due to $\gamma\gamma \rightarrow$ hadrons processes.  

For the conceptual design report of CLIC, a first set of detector concepts \cite{Linssen:2012hp} based on the two ILC concepts ILD \cite{Abe:2010aa} and SiD \cite{Aihara:2009ad} was established. This is motivated by the fact that the general performance requirements for ILC and CLIC are quite similar. For the CLIC detectors, design modifications motivated by the higher collision energy of up to 3 TeV and by the more challenging experimental conditions such as the very high bunch crossing rate of 2 GHz and a higher background rate than at ILC are implemented. Both detector concepts provide highly efficient tracking with excellent momentum resolution in a solenoidal field and precise secondary vertex reconstruction.  Highly segmented calo\-rimeters optimised for jet reconstruction using particle flow algorithms are foreseen. To provide the depth of the calorimeters necessary for multi-TeV operation, the barrel hadronic calorimeter uses tungsten absorbers. The inner radius of the vertex detectors is increased compared to the ILC to account for the larger beam crossing angle and the higher rate of incoherent $e^+e^-$ pairs produced in the collision. The mitigation of the effects of the $\gamma\gamma \rightarrow$ hadrons background requires a time-stamping on the ns to 10 ns level in most detector subsystems.

For the present analysis, the CLIC\_ILD \cite{Linssen:2012hp} detector is used. This detector concept uses a low-mass pixel vertex detector with an innermost radius of 31 mm, intermediate silicon strip tracking and a time projection chamber as main tracker complemented with an outer silicon tracking envelope. The calorimeter system consists of a silicon-tungsten electromagnetic calorimeter and  hadron calorimeters with scintillator tiles read out by silicon photomultipliers with tungsten absorbers in the barrel and steel absorbers in the endcap region. The full detector system is embedded within a 4 T solenoid, with additional muon tracking detectors in the steel return yoke. A detailed model of the detector has been implemented in GEANT4 \cite{Agostinelli:2002hh}, and the full reconstruction software including particle flow reconstruction has been used in the present analysis. 

\section{Simulation and reconstruction}
\label{sec:EventGen} 

\begin{table}
\caption{Squark and neutralino masses and combined right-handed squark production cross section for the CLIC CDR benchmark model at 3 TeV.  \label{tab:masses}}
\centering
\begin{tabular}{cccc}
\hline
$m_{\tilde{u}_R}, m_{\tilde{c}_R}$ & $m_{\tilde{d}_R}, m_{\tilde{s}_R}$ & $m_{\tilde{\chi}}$ & $\sigma_{\mathrm{comb}}$\\
\hline
1125.7 GeV & 1116.1 GeV & 328.3 GeV & 1.47 fb \\
\hline
\end{tabular}

\end{table}

The physical parameters of the process studied here are based on the "SUSY model I" developed for the benchmark studies in the CLIC CDR. The full model details are given in \cite{CLIC_SUSY_Models}. The 1$^{\mathrm{st}}$ and 2$^{\mathrm{nd}}$ generation right-handed squarks in this model decay with a branching ratio of 99.7\% directly to the corresponding quark and the lightest neutralino. The signal process in the present paper is thus
\begin{eqnarray}
e^+e^- \, \rightarrow \, \tilde{q_R}\bar{\tilde{q_R}}\,  \rightarrow \, q\bar{q}\tilde{\chi}^0_1\tilde{\chi}^0_1 . \nonumber
\end{eqnarray}

In the model the 1$^{\mathrm{st}}$ and 2$^{\mathrm{nd}}$ generation right-handed squarks are mass-degenerate, with the up-type squarks 9.6 GeV heavier than the down-type squarks. The relevant particle masses as well as the combined production cross section taking the CLIC luminosity spectrum into account are given in Table \ref{tab:masses}. The cross section for the production of up-type squarks is 3.82 times larger than that of down-type squarks, thus the signal is dominated by up-type squark production.

\begin{table}
\caption{Cross sections of signal and considered background contributions at 3 TeV taking the CLIC 3 TeV luminosity spectrum into account. The background processes are grouped into two classes, based on whether they lead to true missing energy ($E_{miss}$) due to escaping neutrinos or not. \label{tab:backgrounds}}
\centering
\begin{tabular}{lc}
\hline
process & cross section\\

\hline
signal &  \\
$e^+e^- \, \rightarrow \, \tilde{q_R}\bar{\tilde{q_R}}\,  \rightarrow \, q\bar{q}\tilde{\chi}^0_1\tilde{\chi}^0_1 (u,\, d,\, s,\, c) $ & 1.47 fb \\
\hline
SM, no $E_{miss}$ & \\
$ e^+e^- \, \rightarrow \, q\bar{q} $& $\sim$ 3000 fb\\
$ e^+e^- \, \rightarrow \, q\bar{q}e^+e^- $ &  $\sim$ 3300 fb\\
\hline
SM, $E_{miss}$ & \\
$ e^+e^- \, \rightarrow \, q\bar{q} \nu\bar{\nu}$&  $\sim$ 1500 fb\\
$ e^+e^- \, \rightarrow \, q\bar{q} e^\pm  \nu$&  $\sim$ 5300 fb\\
$ e^+e^- \, \rightarrow \, \tau^+\tau^-  \nu\bar{\nu}$&  $\sim$ 130 fb\\
\hline

\end{tabular}

\end{table}

In addition to the signal itself, several potentially important background contributions are considered. They all are characterized by a low multiplicity of highly energetic jets. The cross section of the background processes, compared with the signal, are summarized in Table \ref{tab:backgrounds}. Beyond the dominating Standard Model (SM) processes there are also potential SUSY backgrounds, which have not been considered here. They would have comparable cross sections to the signal, but more complex final states, likely resulting in a substantial rejection by the event selection. Backgrounds of the type $\gamma\gamma \rightarrow\, q\bar{q} (+X)$ and $e^{\pm}\gamma \rightarrow\, q\bar{q} (+X)$  were likewise ignored, since they lead to highly boosted final states and typically do not result in events with very large missing transverse energy ($E_{T}^{miss}$), and are thus expected to be fully rejected by the event selection. Such processes are only considered in the context of beam-induced background as "pile-up" on physics events, as discussed below. After the event selection discussed in Section \ref{sec:selection}, only processes with substantial missing transverse energy remain significant. 

Events, both signal and background, were generated with WHIZARD 1.95 \cite{Kilian:2007gr}, using PYTHIA \cite{Sjostrand:2006za} for the hadronization. The three  dominating background processes, listed under ``SM, $E_{miss}$'' in Table \ref{tab:backgrounds} have cross sections of two to more than three orders of magnitude higher than the signal, resulting in a combined cross section of approximately 8 pb. For an integrated luminosity of 2 ab$^{-1}$, this would result in 16 million events, which is beyond what could be fully simulated for this study. Thus, the number of events that was simulated was reduced by a generator-level cut on the missing transverse energy. Two different thresholds are used, one at $E_{T}^{miss}$ of 530 GeV, above which all events are passed to the full simulation, and one at 330 GeV, above which 10\% of all events which are below the higher threshold are fully simulated. With these cuts, the number of events to be simulated was reduced to approximately 250\,000. To help the signal and background discrimination, a cut on missing transverse energy requiring more than 600 GeV $E_{T}^{miss}$ was applied in the analysis, which means that only events above the higher threshold, which are simulated for the full integrated luminosity, contribute significantly to the final event sample. The background events with intermediate $E_{T}^{miss}$ are used to confirm this expectation, and are not considered in the final analysis. The transverse missing energy rather than the full missing energy is used in this first level of event selection to make it robust against the smearing of collision energies and the resulting boost of the final state system introduced by the luminosity spectrum of the collider. 

For the full simulations, the generated events were passed through a detailed GEANT4 simulation of the CLIC\_ILD detector implemented in the Mokka framework \cite{MoradeFreitas:2002kj}, and were then overlayed with $\gamma\gamma\, \rightarrow\, \mathrm{hadrons}$ events corresponding to 60 bunch crossings using specifically developed software tools \cite{Linssen:2012hp}.

Events were reconstructed using the full CLIC\_ILD reconstruction chain,
including the PandoraPFA particle flow algorithm \cite{Thomson:2009rp} with specific cuts on reconstructed particle
objects \cite{Marshall:2012ry} to reduce the impact of the $\gamma\gamma\, \rightarrow\,
\mathrm{hadrons}$ background. The time stamp of each reconstructed particle is required to match closely with the time of the interesting physics event, where the most powerful timing information is obtained
from the calorimeters. The timing cut varies with the transverse momentum ($p_T$) of the
particle, since low $p_T$ particles are more likely to be background. The cuts reject particles with reconstructed times inconsistent with the reconstructed event time. Particles with  $p_T$ $>$ 4 GeV (8 GeV in the case of neutral hadrons) are accepted irrespective of timing. For particles below that threshold, particle type and region-dependent cuts between 1 ns and 3 ns are applied, with the most severe cuts in the forward region below a $p_T$ of 0.75 GeV. More details are given in
\cite{Linssen:2012hp, Marshall:2012ry}.  In addition to these cuts, jet finding plays a major role in the
rejection of beam related background. Its influence, as well as the impact of the background rejection cuts,
are discussed in the following.

\section{Jet finding}

\begin{figure}
\centering
  \includegraphics[width=0.86\columnwidth]{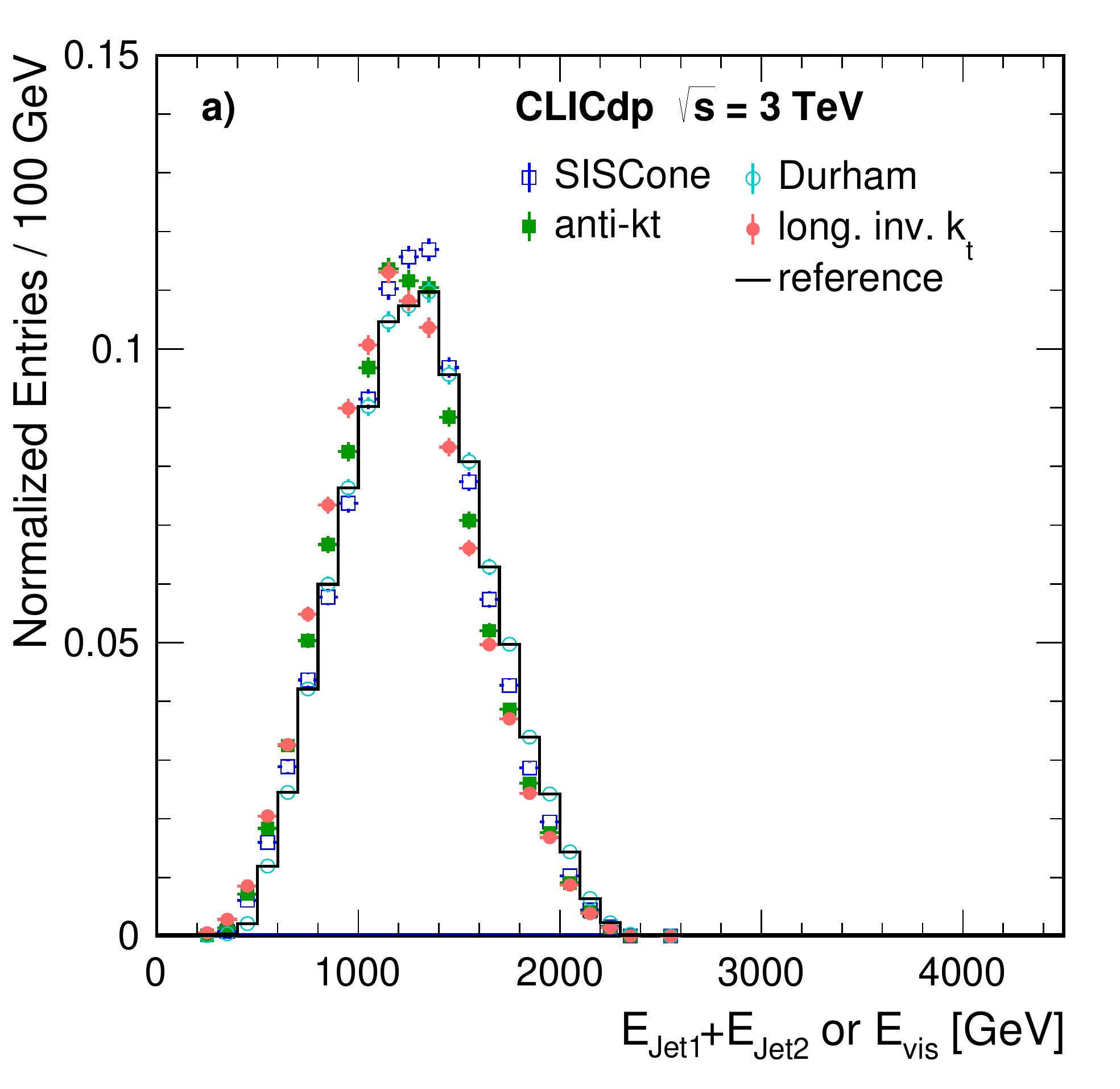}\\
   \includegraphics[width=0.86\columnwidth]{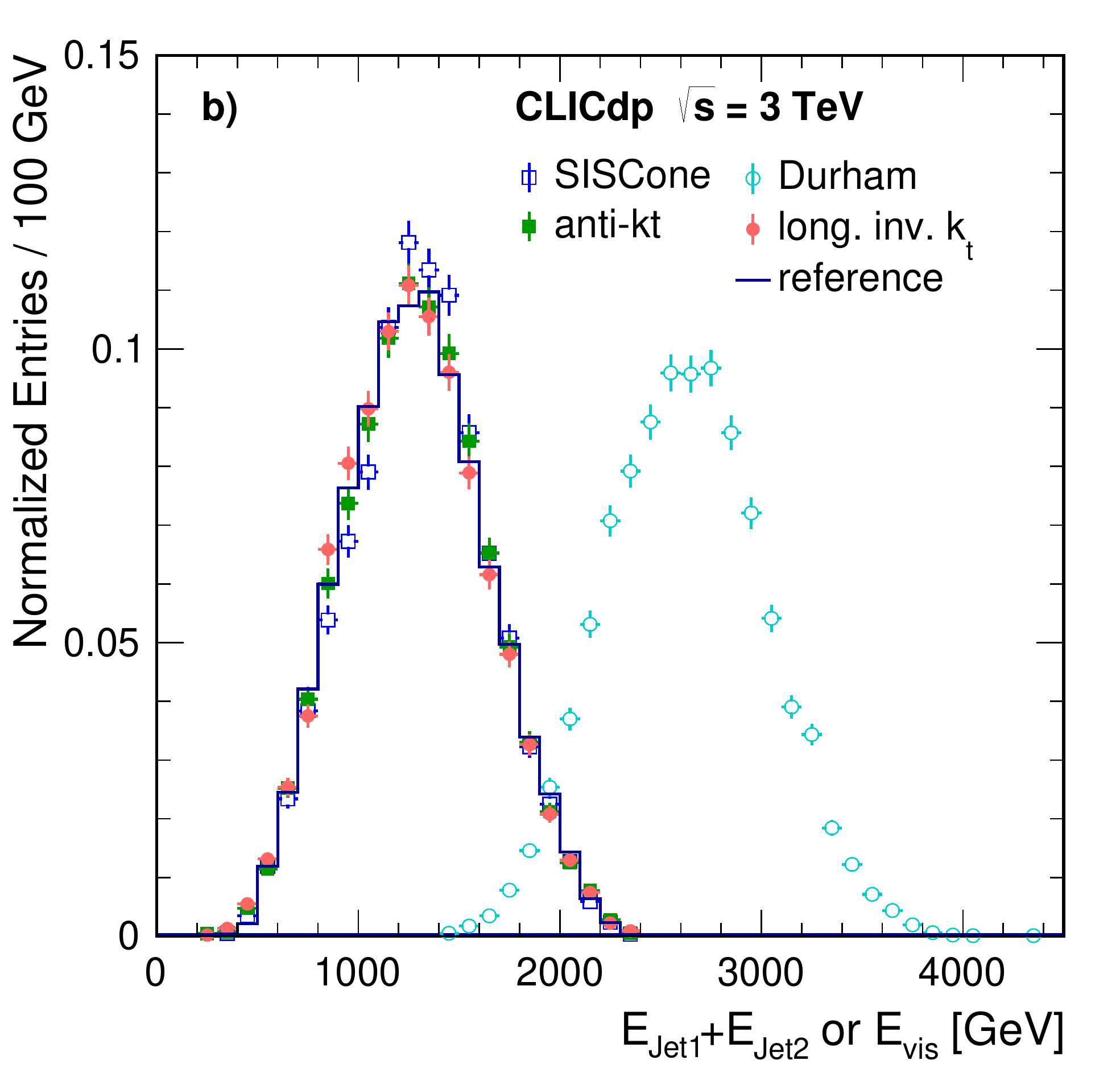}\\
    \includegraphics[width=0.86\columnwidth]{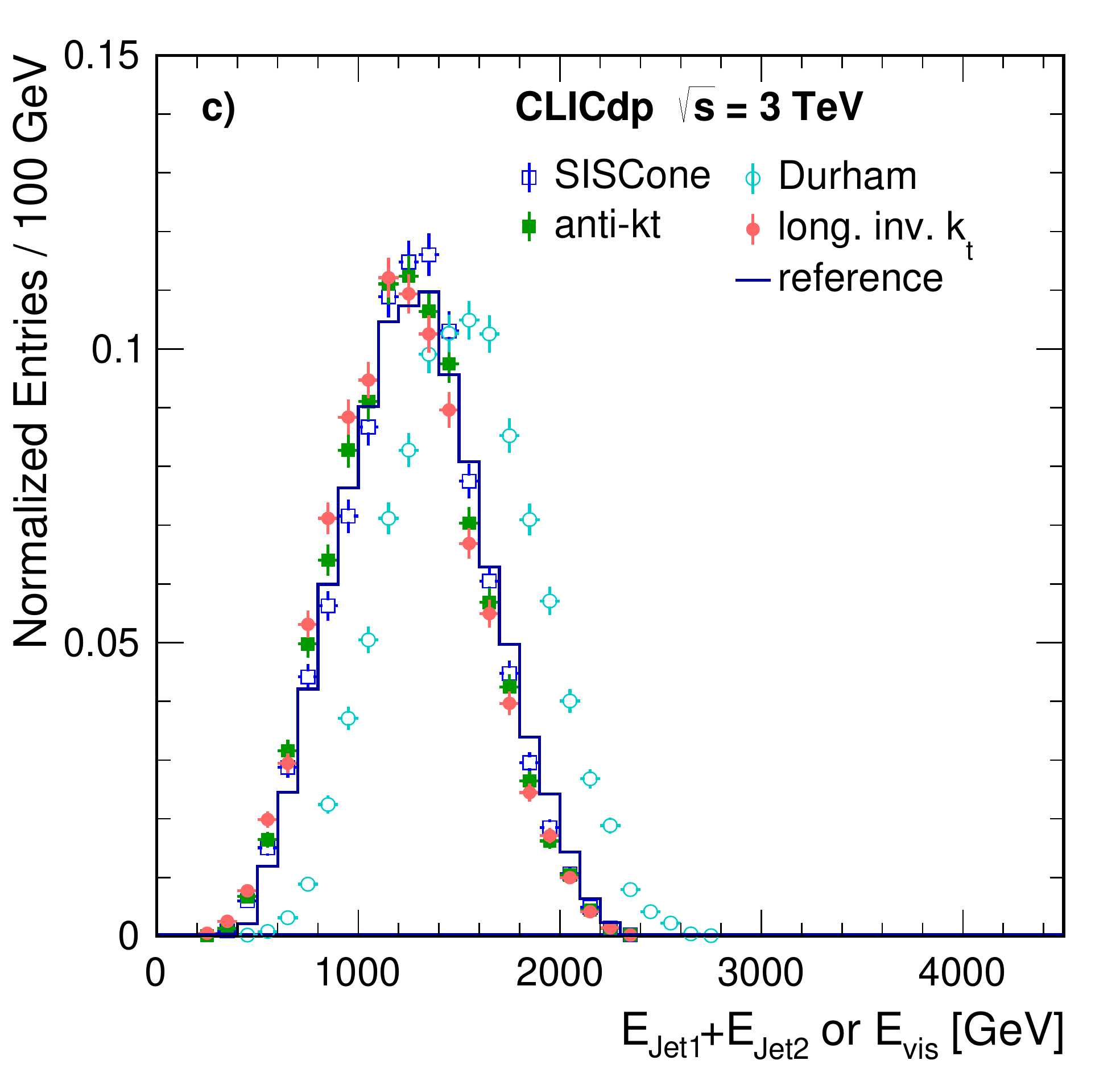}\\
 	\caption{Visible energy, given by the sum of the two highest-energetic jets $E_{\mathrm{Jet1}}$ + $E_{\mathrm{Jet2}}$  for the different jet finding techniques and given by the sum of the energy of all visible particles $E_{\mathrm{vis}}$ for the reference for a) no $\gamma\gamma \rightarrow$ hadrons background without background rejection cuts, b) full background without background rejection cuts and c) full background and the background rejection cuts. }
   \label{fig:JetPerformance}
 \end{figure}

The final state of the signal process considered here, $e^+e^- \, \rightarrow \, \tilde{q_R}\bar{\tilde{q_R}}\,  \rightarrow \, q\bar{q}\tilde{\chi}^0_1\tilde{\chi}^0_1$, contains two neutralinos and two
quarks. While the neutralinos are not visible in the detector, both quarks will
undergo hadronization and each will create a jet. A large variety of different
jet finders for the identification and reconstruction of those jets are available, each with particular strengths and weaknesses. For the
present study, the FastJet library (version 2.4.2) \cite{Cacciari:2005hq, Cacciari:2011ma}, which provides
implementations of various jet algorithms, was used. At CLIC, jet finding is performed on reconstructed particles, the particle flow objects (PFOs) reconstructed by PandoraPFA. For the energy and momentum determination of these objects the information of all sub-detectors, including tacking and calorimetry, is used. 

The key challenge for the event reconstruction at CLIC in general, and for jet finding in particular, is the $\gamma\gamma \to {\rm hadrons}$
background. Since the quality of the jet energy measurement directly enters into the squark mass measurement, it is crucial to limit the amount of background energy erroneously added to the signal jets. At the same time, it is important to avoid significant loss of signal energy outside of the jets used for the mass measurement. Since the background is peaked in the forward region and since the number of jets looked for in the analysis is known, exclusive jet clustering algorithms are a natural choice. These algorithms, originally intended for the environment of hadron colliders, include so-called beam-jets to absorb particles close to the beam axis. These particles originate from projectile remnants in the case of hadron colliders, but the principle can also be applied to forward- and backward-going background at lepton colliders. Exclusive algorithms are also able to cluster a given event into a fixed number of jets, a feature particularly attractive at lepton colliders when analysing specific event signatures. 

The FastJet library provides two different infrared- and collinear-safe exclusive jet algorithms which are variants of the $k_t$ - algorithm optimized for either hadron or electron-positron colliders. The $k_t$ algorithm is a recombination algorithm which recursively clusters particles pairwise based on the distance $d_{ij}$ between two particles $i$ and $j$. The precise definition of the distance measure defines the behavior of the algorithm, and influences its susceptibility to beam-related background. In its original form it
was developed for an electron-positron environment (\texttt{ee\_kt\_algorithm} in FastJet, often referred to as Durham algorithm) \cite{Catani:1991hj}, using the energy $E$ weighted angle $\theta$ between two particles 
\begin{eqnarray}
d_{ij,{\rm ee}} & = & \min(E_i^2, E_j^2) (1-\cos\theta_{ij}) \label{eq:jet:ee-distance}
\end{eqnarray} 
as  distance definition.

At a hadron collider the center-of-mass system has an unknown boost along the beam axis,
leading to the use of only transverse observables. For the $k_t$ - algorithm for
hadron colliders (\texttt{kt\_algorithm}) \cite{Catani:1993hr, Ellis:1993tq} the particle energy is replaced by the
transverse observable $k_t$, and the angle between particles is replaced by the
boost-invariant pseudorapidity $\eta$ and the azimuthal angle $\phi$, resulting
in a longitudinally invariant algorithm with
\begin{eqnarray}
d_{ij,{\rm hadr}} & = & \min(k_{t,i}^2, k_{t,j}^2)\frac{\eta_{ij}^2\phi_{ij}^2}{R^2} \label{eq:jet:kt-distance}
\end{eqnarray}
as two-particle distance definition. Here, $R$ (measured in $\eta$, $\phi$) is a jet size parameter which influences the size of the jets in the event, and for a fixed number of jets also has a strong impact on the size of the beam jets. Since this metric makes use of the pseudorapidity, the distance between particles gets stretched in the forward region, making the algorithm less susceptible to  $\gamma\gamma \to {\rm hadrons}$ pile-up. Both $k_t$ algorithms are considered in the analysis. 

As a comparison, two inclusive jet algorithms have been studied as well, the anti-$k_t$ algorithm \cite{Cacciari:2008gp} and the Seedless Infrared Safe Cone (SISCone) \cite{Salam:2007xv} algorithm. Those two algorithms use the same metric as the longitudinally invariant $k_t$ algorithm, and are thus also expected to be rather robust against beam-induced background.

All of the considered algorithms, with the exception of the Durham algorithm, have one tuneable parameter, the jet radius parameter $R$. For the anti-$k_t$ and the SISCone algorithms, this parameter influences the typical size of the jets, and with that the number of jets that are found in a given signal event. For the exclusive, longitudinally invariant $k_t$ algorithm, the $R$ parameter influences the amount of particles that are clustered into the beam jets, and with that the total visible energy that is included in the pre-defined number of reconstructed jets. In this study, the $R$ parameter is tuned on events with background to obtain a good agreement of the visible energy in the case of the $k_t$ algorithm, and to achieve a clustering of the majority of the signal events into two jets for the anti-$k_t$ and the SISCone algorithms. For the $k_t$ algorithm, $R = 0.7$ is chosen, while $R= 0.5$  is found to be optimal for the anti-$k_t$ and the SISCone algorithm.

To compare the performance of the different jet finding algorithms, two different scenarios for timing and momentum cuts applied in the event reconstruction are considered. The sensitivity of the different algorithms to $\gamma\gamma \to {\rm hadrons}$ background is illustrated on events without timing cuts. In those events, the added 60 bunch crossings of background contribute an excess energy of 1.4 TeV in the main calorimeters. The performance in a realistic environment is studied using timing cuts in the event reconstruction as discussed in Section \ref{sec:EventGen}. In addition, a baseline comparison is provided with events without the overlay of background.

Figure \ref{fig:JetPerformance} shows the sum of energy of the two most energetic jets found in the event for squark signal events for different jet finders, compared to the reference distribution of the true signal-only energy, given by the full visible energy E$_{\mathrm{vis}}$ from the signal MC event record for different background scenarios. Figure \ref{fig:JetPerformance} a) shows the situation without the overlay of $\gamma\gamma \to {\rm hadrons}$ events, Figure \ref{fig:JetPerformance} b) the case for a full 60 bunch crossings of background without any timing cuts, and Figure \ref{fig:JetPerformance} c) the realistic case with the application of timing and transverse momentum cuts on the reconstructed particle flow objects. While the classic Durham algorithm provides the best agreement with the reference distribution in the case of no background, it picks up almost the complete 1.4 TeV of additional energy when the full background is included, and also severely suffers from background pickup when timing cuts are used. This makes the algorithm unsuitable for jet finding at CLIC at energies where the contributions from $\gamma\gamma \to {\rm hadrons}$ processes are sizeable, which is the case for energies of 500 GeV and above. The other, hadron-collider-inspired, algorithms all show a comparable performance when reconstructing the visible energy. For analyses that target specific final-state signatures such as the present squark analysis, the possibility of forcing the events to be clustered into a pre-defined number of jets provided by exclusive algorithms is very convenient, leading to the choice of the longitudinally invariant $k_t$ algorithm for the physics benchmark analyses at CLIC \cite{Linssen:2012hp}. While this procedure favors the inclusion of additional particles into the signal jets which otherwise are reconstructed as additional jets, it is found that this still results in a better definition of the kinematic observable used in the squark analysis, discussed in detail below, than the use of inclusive algorithms with variable number of jets.  


Recently, a new algorithm, the Valenica algorithm, has been introduced, based on the experience with jet finding in the CLIC and ILC physics studies \cite{Boronat:2014hva}. This algorithm combines features of the longitudinally invariant $k_t$ algorithm and the classic $k_t$ algorithm. Since this algorithm was proposed after the completion of the physics studies for the CLIC CDR it is not considered in the present analysis.

\section{Mass measurement techniques}

The final state of the considered process $
e^+e^- \, \rightarrow \, \tilde{q_R}\bar{\tilde{q_R}}\,  \rightarrow \, q\bar{q}\tilde{\chi}^0_1\tilde{\chi}^0_1$ 
is characterized by two highly energetic jets and missing energy. The mass of the squarks is extracted from the measured jets. Several different techniques for this extraction have been evaluated in the course of this study \cite{LCD-2010-012}. From the end-points of the energy distribution of the final-state jets, the mass of both the squark and the $\tilde{\chi}^0_1$ can be extracted \cite{Feng:1993sd}. The upper edge of this distribution is substantially distorted by the luminosity spectrum of the collider. Recent studies have shown that the luminosity spectrum at CLIC can be measured with sufficient precision to still enable a precise determination of the edge position \cite{Poss:2013oea}. However, the jet energy distribution, in particular the lower edge given by low-energetic jets, also suffers significantly from Standard Model background, making precision measurements challenging. 

Here, a different technique is explored. It is assumed that the mass of the lightest neutralino will be measured with satisfactory precision in processes with higher cross sections and less background sensitivity, such as slepton production and decay \cite{Battaglia:2013bha}. With this additional knowledge, the extraction of the squark mass from distributions with a single kinematic edge becomes possible. 

Since the distribution of the center-of-mass energy at a 3 TeV CLIC collider has a substantial tail towards lower energies due to beamstrahlung, with only 35\% of the luminosity in the top 1\% of the energy, methods which do not rely on the knowledge of the precise center-of-mass energy are advantageous. One such technique is the variable $M_C$  \cite{Tovey:2008ui}, which uses the momenta of the two observed jets to form a modified invariant mass which is invariant under contra-linear boosts of  equal magnitude of the two squarks, and thus independent of the center-of-mass energy. $M_C$ is given by
\begin{eqnarray}
M_C &=& \sqrt{(E_{q,1} + E_{q,2})^2 - (\vec{p}_{q,1} - \vec{p}_{q,2})^2}\\
&=& \sqrt{2 (E_{1}E_{2} + \vec{p}_{1} \cdot \vec{p}_{2})},
\end{eqnarray}
where $E_{1}$, $\vec{p}_{1}$ and $E_{2}$, $\vec{p}_{2}$ are the energies and
three momenta of the two visible final-state quarks (jets), respectively. The
distribution of $M_C$, has a triangular shape with the maximum at the upper edge, given by 
\begin{equation}
M_C^{max} = \frac{m_{\tilde{q}}^2 - m_{\tilde{\chi}^2}}{m_{\tilde{q}}},
\end{equation}
where $m_{\tilde{q}}$ is the mass of the squark and $m_{\tilde{\chi}^2}$ is the mass of the lightest neutralino.
The squark mass is thus given by
\begin{equation}
 m_{\tilde{q}} = \frac{1}{2} \left( M_C^{max} + \sqrt{(M_C^{max})^2 + 4 m_{\tilde{\chi}}^2}\right), \label{eq:MassfromMC}
\end{equation}
which can be determined by the measurement of the position of the upper edge of the $M_C$ distribution alone, assuming that the neutralino mass is known from other measurements. In practice, the shape of the edge is slightly smeared out by detector resolution effects, which need to be accounted for in this analysis.

The construction of $M_C$ assumes that the center-of-mass system of the
collision is at rest in the detector system, which evidently is not the case at
CLIC due to beamstrahlung and initial state radiation. Still, the boost of the
collision system with respect to the laboratory frame is typically quite small for the high center-of-mass energies required to produce TeV-scale objects,
making it advantageous to use the complete available information, and not just
transverse observables, as would be done at hadron colliders. The beam energy
spectrum leads to a slight distortion of the edge of the $M_C$ distribution, which can be precisely modelled based on the knowledge of the luminosity spectrum. The best precision on the squark mass can thus be obtained from template fits which take the effect of the beam
energy spectrum of CLIC into account, allows the inclusion of effects of selection cuts on the $M_C$ distribution and maximizes the use of the available statistics by considering all reconstructed events. 

\section{Squark identification and mass measurement}

The analysis itself is divided into two separate steps: The identification of candidate signal events in an environment with high Standard Model backgrounds, and the determination of the squark mass from the identified signal candidates. 

\subsection{Signal selection}
\label{sec:selection}

\begin{figure*}
\centering
  \includegraphics[width=0.86\columnwidth]{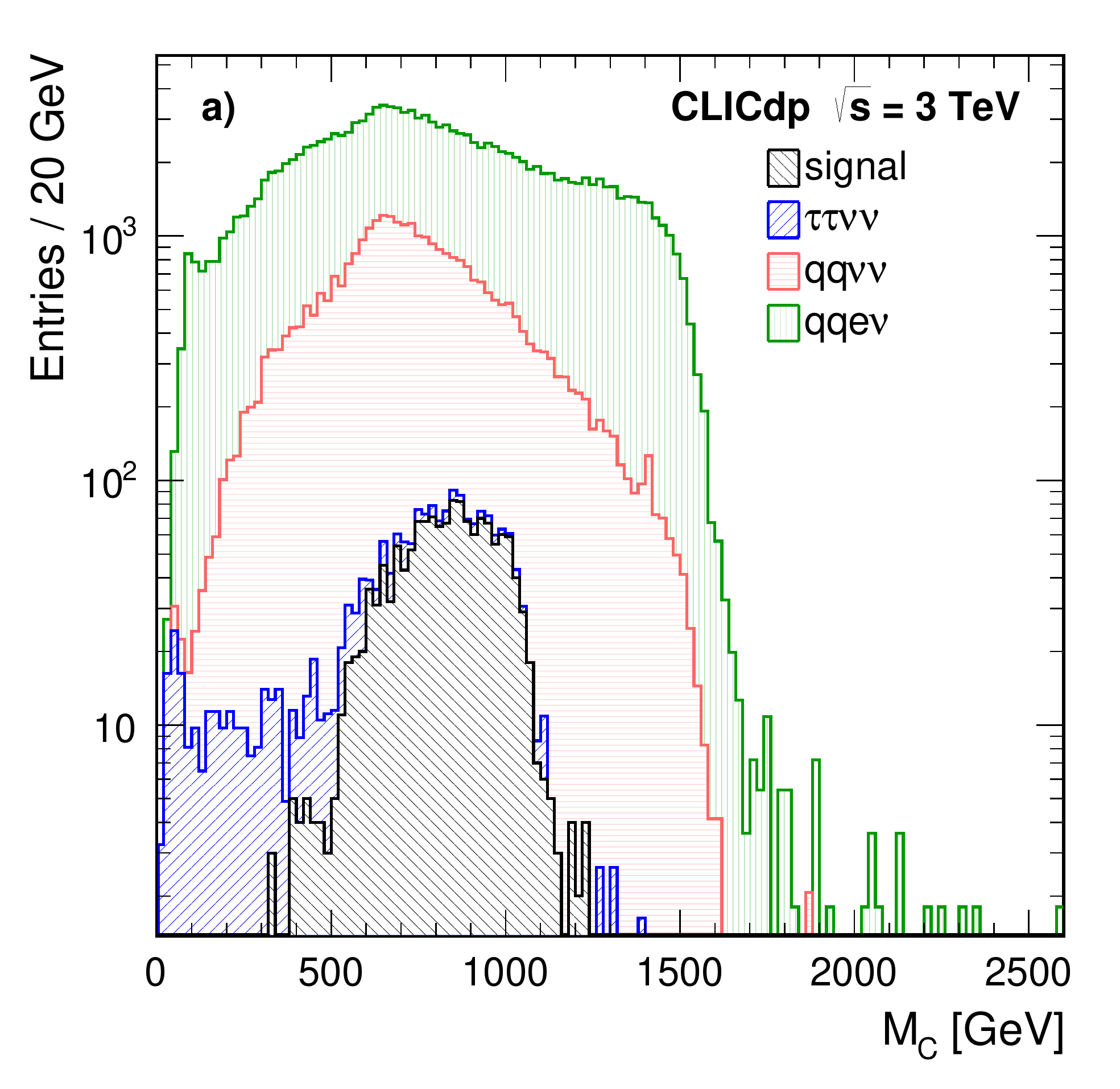}
  \includegraphics[width=0.86\columnwidth]{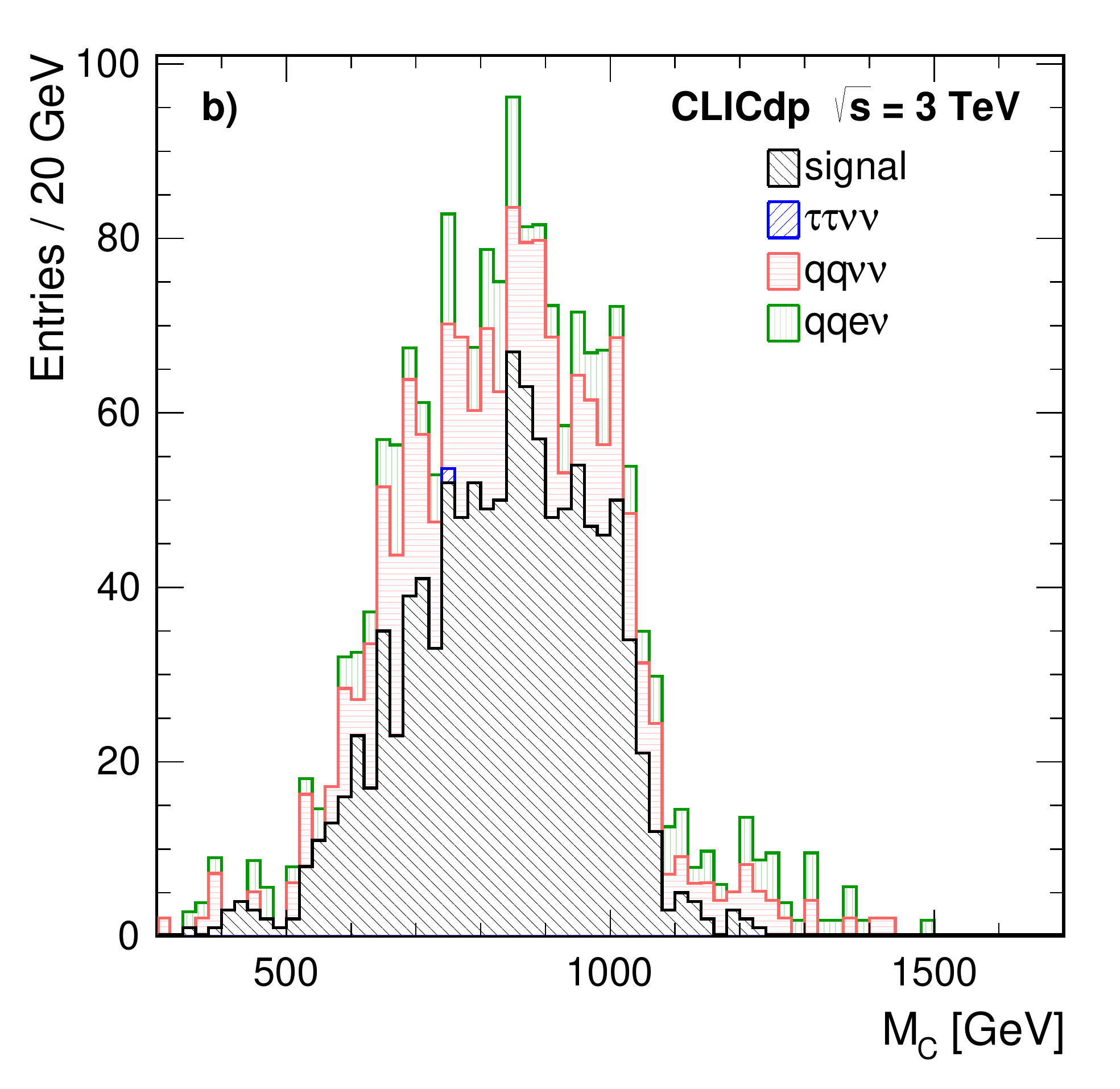}\\
  \includegraphics[width=0.86\columnwidth]{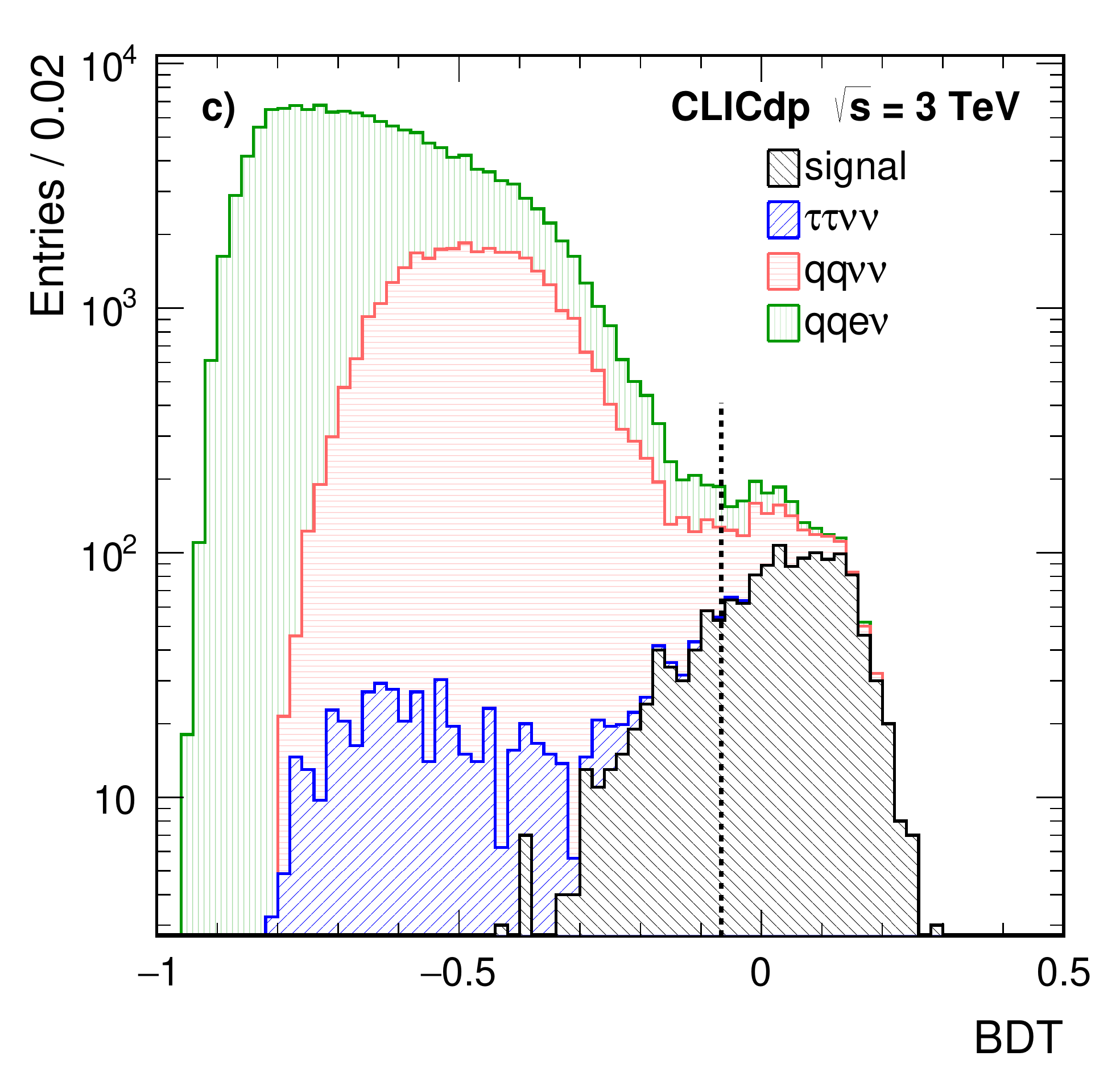}  
  \includegraphics[width=0.86\columnwidth]{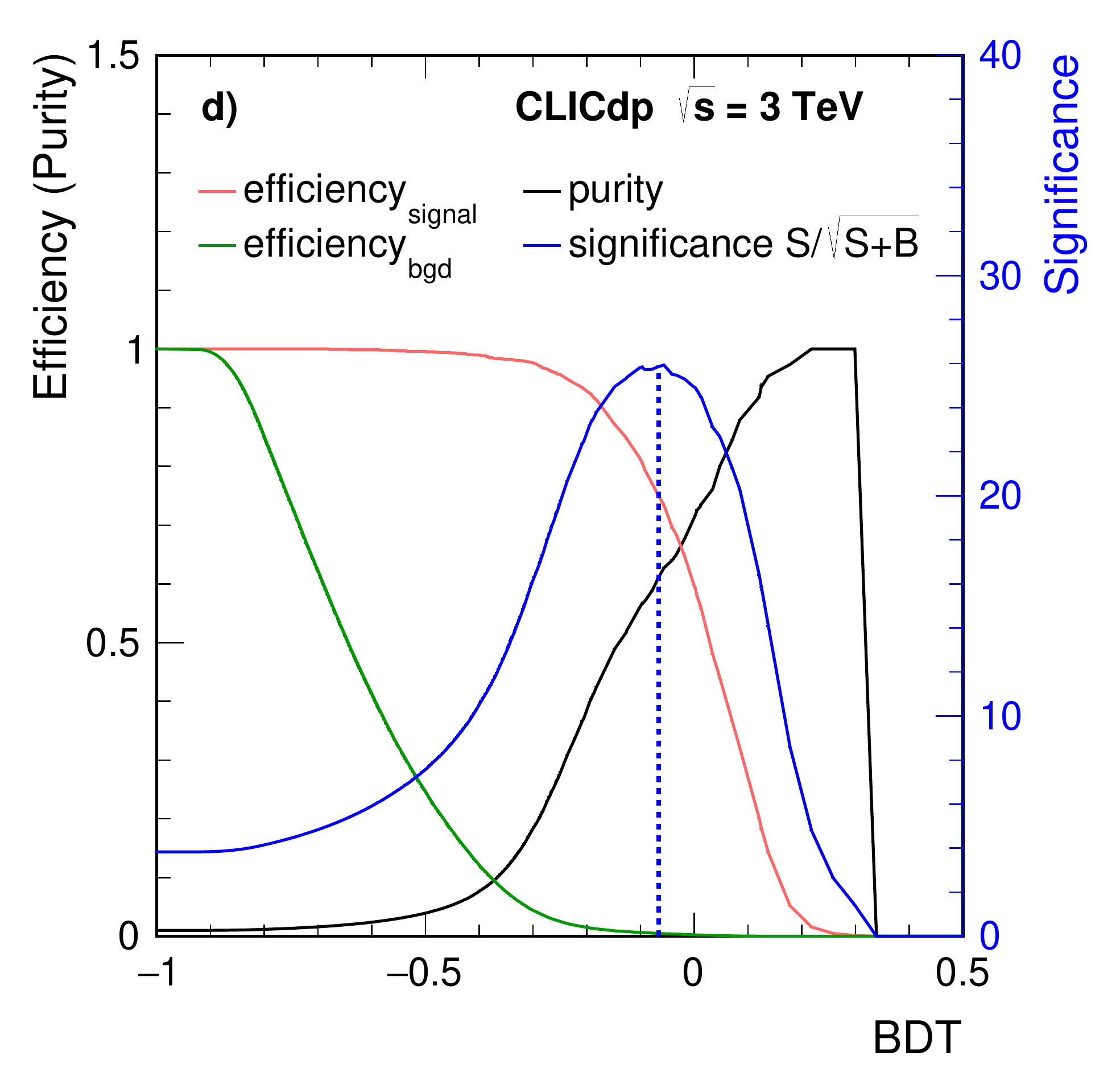}
  
 \caption{Signal selection based on boosted decision trees. a) Stacked $M_C$  distribution of the signal and the three main backgrounds after the application of a cut requiring  $E_{T}^{miss}$ > 600 GeV.  b)  Stacked $M_C$  distribution of the signal and the three main backgrounds after the application of the cut on the BDT output. c) Stacked histrogram of the distribution of the BDT classifier for the signal and the three main backgrounds, with the vertical line indicating the cut at -0.066 applied in the event selection to achieve the highest signal significance. d) Efficiencies for signal and background (bgd), signal purity and significance as a function of the cut on the multivariate classifier $BDT$, with the vertical line indicating the point of highest signal significance at -0.066. All histograms are normalized to an integrated luminosity of 2 ab$^{-1}$.}
   \label{fig:SquarkSelection}
 \end{figure*}

The rejection of non-squark physics background is one of the main challenges in
the present analysis. Since the signal signature of two jets and missing energy
is rather generic, high cross section Standard Model processes contribute to the
background. As discussed above in Section \ref{sec:EventGen} and summarized in
Table \ref{tab:backgrounds}, three major background channels are
considered in the final analysis: $\tau^+\tau^-\nu\bar{\nu}$, $q\bar{q}\nu\bar{\nu}$ and $q\bar{q}e^{\pm}\nu$. All three have  cross
sections which are two to three orders of magnitude above the cross section of
light-flavor right-handed squark production.
 
 A high signal purity, in particular in the region of the kinematic edge of the
 distribution, is crucial to obtain a precise mass measurement. This requires a
 reduction of the background by more than a factor of 1000. A cut on missing
 transverse momentum provides a first substantial reduction of the background level. Requiring a measured  $E_{T}^{miss}\, >\, 600$ GeV
 reduces the dominating background channels to approximately $10^{-2}$, in the
 case of the  $\tau^+\tau^-\nu\bar{\nu}$ final state even to $2\,\times\,10^{-3}$, of their
 original cross section, while reducing the signal to 0.485 of the original sample.
 
This reduction alone is insufficient, as illustrated in Figure 
\ref{fig:SquarkSelection} a), where the background still dominates the $M_C$ -
distribution over the full kinematic range. Hence additional background
rejection is necessary.

A further reduction of the Standard Model background is achieved  by
using a forest of 150 Boosted Decision Trees (BDT) implemented in the ``Toolkit for
Multivariate Data Analysis'' \cite{Hocker:2007ht} for ROOT \cite{Brun:1997pa}. The training and testing of the BDT is performed on dedicated data samples.
The 150 decision trees are boosted using adaptive boosting. No pruning is performed. Instead, each tree
is allowed to have a maximum depth of 3 levels. The multivariate analysis uses a total of 18 event variables. These are the energy, the invariant mass, the angle with respect to the beam axis and the type and energy of the leading particle of each of the two jets, the acoplanarity of the two jets, the ratio of the number of particles in the two jets, the total number of leptons and particles in the event, the energy of the most energetic lepton in the event, the total $E_{T}^{miss}$ after jet finding and the distance parameters $y_{12}$ and $y_{23}$ at which the jet finder would find one instead of two and three instead of two jets. 


In the analysis, each tree evaluates the event and classifies it as either signal (+1) or
background (-1). The (weighted) average of all 150 trees creates a single
response variable referred to as $BDT$ in the following. This variable provides a clear separation of signal and background, as shown in Figure \ref{fig:SquarkSelection} c). In particular the channel  $e^+e^- \to \tau^+\tau^-\nu\bar{\nu}$ is rejected completely, primarily due to very different jet properties in $\tau$ decays. Depending on the exact choice of the cut on $BDT$, large fractions of the other two backgrounds, which can result in a more signal-like topology, are rejected as well. Further details on the event selection procedure are given in \cite{WeustePhD}.

The optimal cut value was determined from the distributions of purity,
efficiency and corresponding signal significance $S/\sqrt{S+B}$, shown in Figure
\ref{fig:SquarkSelection} d), in order to achieve the highest significance. The highest significance,  $S/\sqrt{S+B} = 25.8$
was achieved with a cut of $BDT >
-0.066$, indicated by the dashed black line in Figure \ref{fig:SquarkSelection} c). With this cut, the total signal efficiency is $\epsilon\, =\, 0.362$ including the cut on missing transverse momentum prior to the application of the BDT, and the obtained overall purity is 0.613. 
The effect of the background rejection on the $M_C$ distribution using the boosted decision trees and the $E_{T}^{miss}$ cut is shown in Figure
\ref{fig:SquarkSelection} b), which shows the stacked distribution of the signal and the three considered background processes after the application of all event selection cuts. 

\begin{figure}
\centering
  \includegraphics[width=0.86\columnwidth]{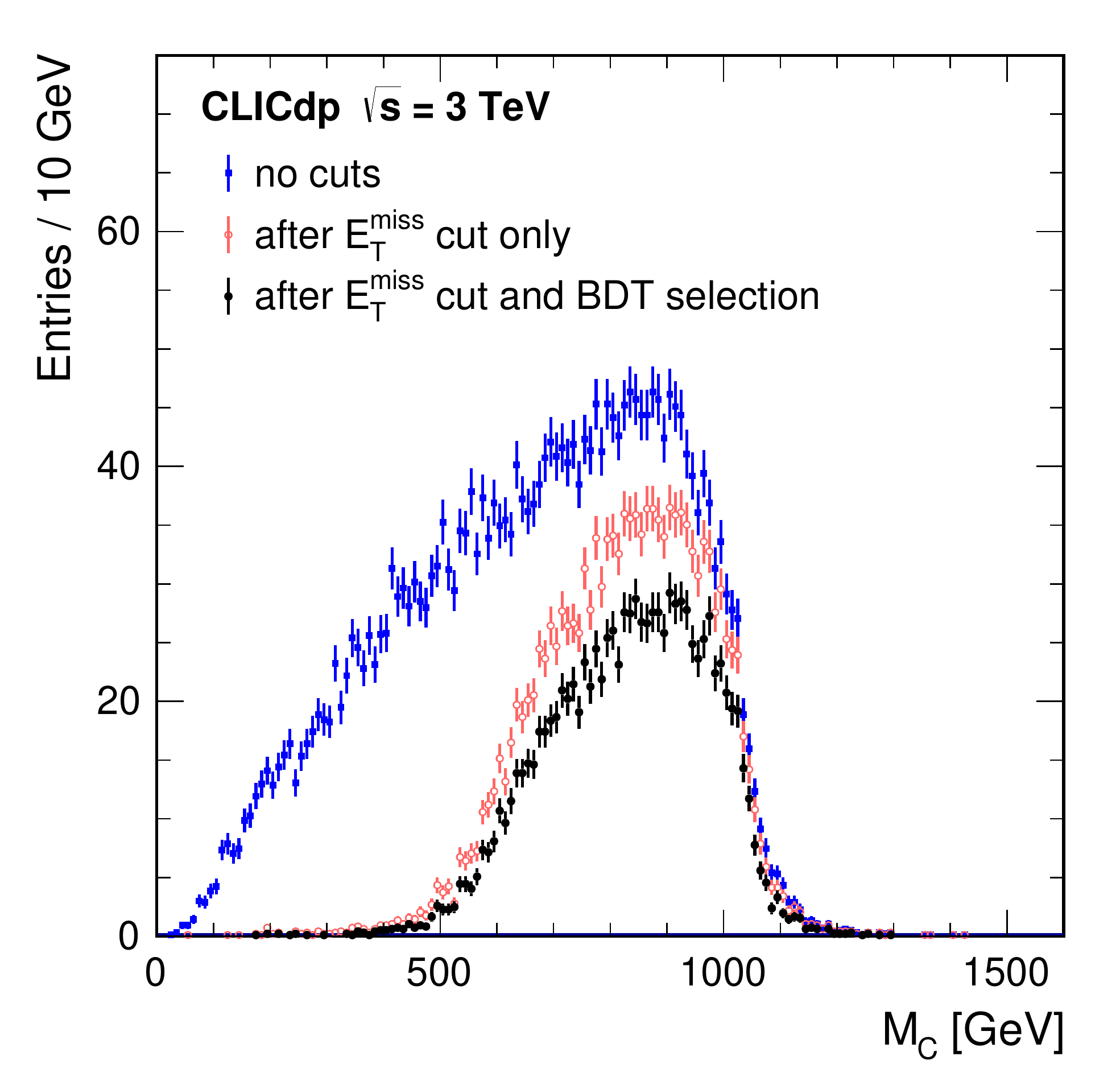}
 \caption{$M_C$ distribution for all signal events in a high-statistics signal only sample, and those selected by the event selection procedure with the $E_{T}^{miss}$ cut alone and with the full selection including the BDT. The normalization of the histograms is scaled to an integrated luminosity of 2 ab$^{-1}$. The shape of the upper edge, which is most sensitive to the squark mass, is not substantially affected by the event selection.}
   \label{fig:SelectionEfficiency}
 \end{figure}

Figure \ref{fig:SelectionEfficiency} illustrates the influence of the signal selection ($E_{T}^{miss}$ cut and BDT) on the $M_C$ distribution, by showing a high-statistics signal-only sample prior to the event selection, after the $E_{T}^{miss}$ cut alone, and after the full event selection. The event selection does not affect the shape of the upper edge, apart from a reduction in overall signal amplitude. The lower $M_C$ region is affected by the $E_{T}^{miss}$ cut, but not further modified by the application of the BDT. This shows that the event selection itself does not introduce a sizeable bias in the mass measurement. Potential systematics arising from the event selection procedure are discussed in more detail in Section \ref{sec:Systematics}.

Cross-checks of the multivariate classifier distributions for signal and background 
performed by comparing the distributions for the training sample with those for the
testing sample show that the classification works successfully. For the
chosen cut value of $BDT> -0.066$, a signal significance of
$S/\sqrt{S+B}\, = \,25.9$ is observed for the training sample, while
$S/\sqrt{S+B}\, =\, 25.7$ is achieved for the testing sample. The good
agreement of these two values shows that possible overtraining effects are
negligible.

\begin{figure}
\centering
  \includegraphics[width=0.86\columnwidth]{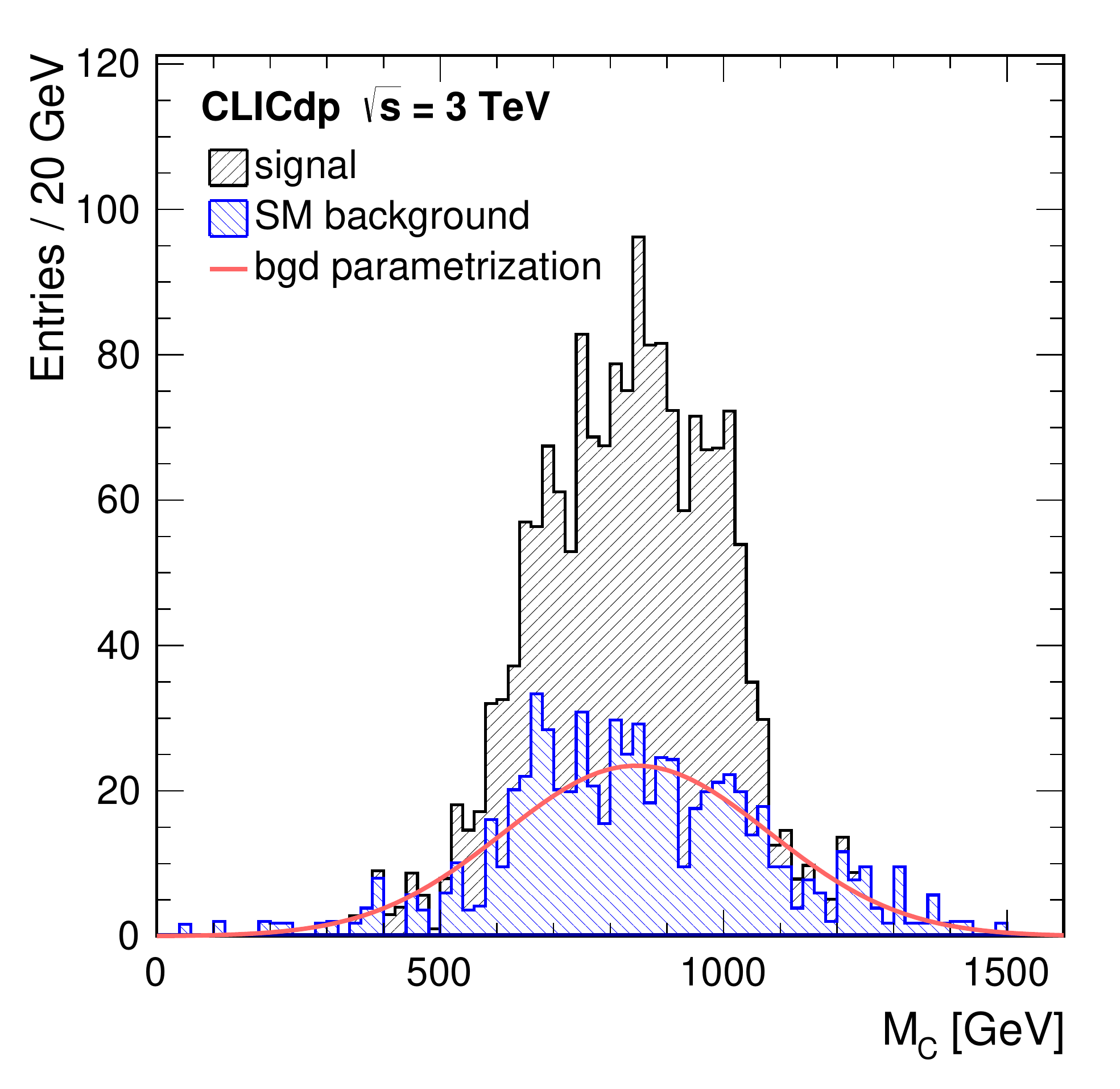}
 \caption{Signal and background distributions after the event selection, shown by the stacked histograms. The parametrization of the background (bgd) with a Gaussian, used for the background subtraction in the mass extraction, is shown by the solid line. The event statistics correspond to an integrated luminosity of 2 ab$^{-1}$.}
   \label{fig:BackgroundParametrization}
 \end{figure}

The mass measurement, discussed below, is performed on a background-subtracted distribution. The background distribution is taken into account by parametrising the  $M_C$ distribution of surviving background events after the event selection by a simple Gaussian function, as illustrated in Figure \ref{fig:BackgroundParametrization}. This parametrization is then used to subtract the background from the $M_C$ distribution obtained from the analysis sample prior to the actual squark mass extraction.

\subsection{Mass measurement}

It is possible to extract the mass of the squarks using the upper edge $M_C^{max}$ of the $M_C$ distribution. For a reliable fit of the edge, the detector resolution, distortions due to the beam energy spectrum and the influence of beam-induced backgrounds need to be accounted for in the fit function. Still, even small background contributions in the region of the edge can have a significant influence, potentially resulting in biased results. 

It thus seems advantageous to use a template fit instead, which allows the inclusion of the above-mentioned effects in the generation of the templates. In such a fit, the mass is determined by comparing the observed distribution with high-statistics signal templates generated for various different squark masses. An additional advantage of a template fit is that the complete $M_C$ distribution enters into the fit, not just the high-$M_C$ edge (although the edge also is the driving region in a template fit), potentially leading to reduced statistical errors for statistically limited samples and resulting in higher stability against remaining background contributions and statistical fluctuations.

\subsubsection{Template generation}

In the model used for this analysis the up-type right-handed squarks
($\tilde{u}_R, \tilde{c}_R$) with mass $m_{\tilde{u}}$ are 9.6 GeV
heavier than their down-type counterparts ($\tilde{d}_R, \tilde{s}_R$)
with mass $m_{\tilde{d}}$.

The present analysis is unable to distinguish between up and down type
squarks. Hence the mass measurement will give the result of the mean squark mass denoted by
$m_{<\tilde{q}>}$, weighted with the respective production cross sections

\begin{eqnarray}
\frac{\sigma_{\tilde{u}}}{\sigma_{\tilde{d}}} & = & 3.82 \\
m_{<\tilde{q}>} & = & \frac{ \sigma_{\tilde{u}} m_{\tilde{u}} + \sigma_{\tilde{d}} m_{\tilde{d}} } {\sigma_{\tilde{u}} + \sigma_{\tilde{d}}}, 
\end{eqnarray}
and is thus dominated by the up-type mass.

For the template generation, a similar mass splitting between up- and down-type
squarks of exactly 10 GeV is used. The templates are created in steps of
3 GeV over the range of 1050 GeV $\leq$ $m_{\tilde{u}}$ $\leq$ 1248 GeV.

In order to minimize statistical fluctuations in the templates, each of these
mass points is generated with 50\,000 events, corresponding to an integrated luminosity
of ${\cal L}$ = 33.6 ab$^{-1}$ at the true squark mass. The templates do not include overlayed beam-induced background since the usage of the $k_t$ algorithm, in particular in combination with the background rejection cuts of the particle flow algorithm, reduces its impact on the $M_C$ distribution to a negligible level. 

Due to computational limitations it was not possible to perform a full
simulation and reconstruction of these 3.3 million events. Instead, detector effects are included on generator level.

As a first step, acceptance is taken into account by rejecting particles with 
$| \cos \theta | > 0.995$ or $p < $ 100 MeV. Then, jet clustering is performed using  the same algorithm as used
elsewhere in this analysis ($k_t$ algorithm with $R=0.7$).
To account for detector resolution effects, the reconstructed jet energies were then smeared with a Gaussian with a width of 4.5\%, obtained by comparing  the shape of the 
$M_C$ - distribution of the smeared jets with a full simulation for one mass point. This smearing is compatible with the performance of PandoraPFA, with a jet energy resolution of approximately 3.5\% to 4\%
${\rm RMS}_{90}$ (the rms in the smallest range of reconstructed energy which contains 90\% of the events) for TeV-scale jets. 

This simplified procedure of including detector effects only acts on the jet energy, but does not affect other details such as particle number and particle identification, which enter the BDT-based background discrimination. The application of the BDT to the smeared generator-level templates would thus not result in the same behavior as for the fully simulated events. Since the BDT does not have a significant influence on the shape of the $M_C$ distribution after the $E_{T}^{miss}$ cut as shown in Figure \ref{fig:SelectionEfficiency}, it is not applied for the generation of the templates. Possible systematic effects due to the omission of this selection are substantially smaller than the statistical uncertainties, and are discussed in Section \ref{sec:Systematics}. The  $E_{T}^{miss}$ requirement on the other hand substantially changes the low-$M_C$ part of the distribution and is included. The templates do thus reproduce the mass-dependent shape of the distribution, but not the reconstruction efficiencies. The overall normalization of the templates is thus kept as a free parameter in the fitting procedure. The templates can be used to extract the mass, but an extraction of the measured cross section requires the use of the signal selection efficiency determined from fully simulated events.

\subsubsection{Template fit}

\begin{figure}
\centering
  \includegraphics[width=0.86\columnwidth]{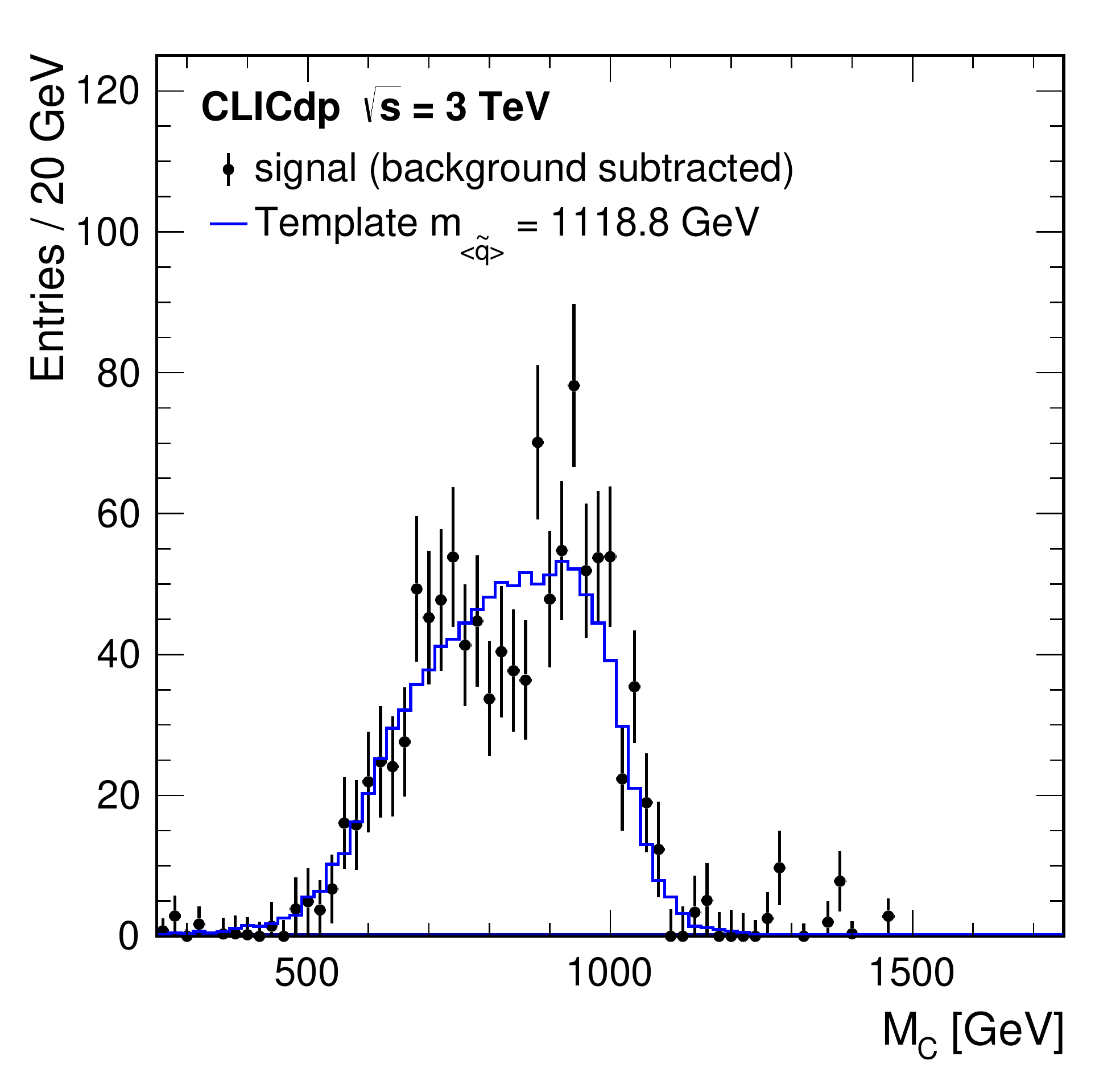}
 \caption{Comparison of the template with the lowest $\chi^2$ to the simulated data points after subtraction of the parametrized background distribution.}
   \label{fig:Template}
 \end{figure}

The template fit is performed by comparing the $M_C$ distributions of different
templates and the background-subtracted simulated measurement corresponding to an integrated luminosity of 2 ab$^{-1}$ using a binned $\chi^2$,
given by
\begin{eqnarray*} 
\chi^2 = \sum_n^{\rm bins} \frac{\Delta_{n}^2}{\sigma_{m,n}^2 + \sigma_{t,n}^2},
\end{eqnarray*}
with $\Delta_n$ giving the difference between measurement and template in bin $n$
and $\sigma_{m,n}$ ($\sigma_{t,n}$) being the statistical error of bin $n$ for
the measurement and the template, respectively. Since the normalization of the templates is not defined with respect to the simulated data due to the absence of the BDT selection in their creation, an individual minimization of the $\chi^2$
was performed for each template, with the template scale factor as free parameter. Figure \ref{fig:Template} shows the $M_C$ distribution of the background-subtracted measurement and that of the template with the lowest $\chi^2$, showing the good match between the distributions. 

The measured squark mass is given by the minimum of the $\chi^2$ distribution as
a function of squark mass. The distribution follows the expected parabolic shape around the minimum, which is determined with a fit of a parabolic function. 

The statistical uncertainty of the squark mass measurement is obtained by performing toy-MC experiments. Each experiment takes the histogram of the MC distribution for the measurement as input and randomly shifts the value of each bin of the histogram according to a Gaussian with a standard deviation corresponding to the statistical uncertainty of this bin. Then the squark mass extraction as described above is repeated. From the width of the distribution of the squark masses determined in 500 such experiments, the statistical uncertainty of the measurement is determined. 

The extraction of the mass from the simulated measurement corresponding to an integrated luminosity of 2 ab$^{-1}$ results in a measured squark mass of 
\begin{eqnarray*}
m_{<\tilde{q}>}  =  1125.5\ \mathrm{GeV}\, \pm\, 6.5\ \mathrm{GeV}\, \mathrm{(stat)}.
\end{eqnarray*}
This corresponds to a relative uncertainty of 0.58\% and is in excellent agreement with the input value of 1123.7 GeV. 

\subsection{Systematic uncertainties}
\label{sec:Systematics}

A full study of all possible systematic uncertainties has not been performed, but several key aspects have been evaluated. 

Since the mass measurement technique used here requires the mass of the neutralino as external input, the uncertainty on this mass assumed to be known from other measurements enters as a systematic on the squark mass. For the particle masses considered here, an uncertainty of 1 GeV on the neutralino mass translates to an uncertainty of 0.54 GeV in the squark mass. From slepton production, uncertainties on the level of 3.4 GeV are expected \cite{Battaglia:2013bha}, which would result in a systematic uncertainty of 1.8 GeV. 

The reconstruction of $M_C$ is based entirely on the reconstruction of jets and is thus very sensitive to the jet energy scale. The position of the upper edge of the distribution, which is most sensitive to the squark mass, shifts as $\Delta j /\sqrt{2}$, where $\Delta j$ is the relative shift of the jet energy from the true scale. For the particle masses in the concrete analysis this translates to an uncertainty of $0.64\,\cdot\,\Delta j$ on the reconstructed squark mass, which corresponds to a mass uncertainty of 7.2 GeV for a jet energy scale uncertainty of 1\%. 

On the analysis side, three potential sources for systematic uncertainties are investigated. The mass extraction technique with  generator-level templates is tested for possible biases introduced by potential systematic differences between the templates without the multivariate event selection and the fully simulated distribution. For this test, the mass measurement is applied to a high-statistics signal-only sample. No statistically significant deviation from the input mass value is observed, excluding mass biases larger than 1.2 GeV.  In addition, the BDT, which is trained for a specific squark mass, can result in systematic shifts due to the choice of this training value. To investigate this bias, the BDT training was repeated with fully simulated data samples with positive and negative mass offsets of 10 GeV and 20 GeV. This study has shown an 80 MeV shift of the reconstructed mass per GeV offset in the BDT training sample. The bias originating from the event selection procedure can thus be reduced to a level substantially below the statistical uncertainty in an iterative procedure. Also the subtraction of non-squark background prior to the template fit is a potential source for systematic uncertainties. To investigate the sensitivity of the mass measurement to this subtraction, the normalization of the background parametrization is varied in the template fit. A $\pm$\,10\% variation of the normalization, corresponding to a 10\% uncertainty in the background efficiency, results in a shift of $\mp$\,0.5 GeV in the reconstructed mass. Since the standard-model background does not exhibit the same sharp upper edge as the signal, the mass measurement is not very sensitive to uncertainties in the background subtraction.

Another source of systematics at CLIC is the knowledge of the luminosity spectrum of the collider, which enters into the mass reconstruction. Here, a simplified, conservative approach has been used by investigating two modified spectra where 5\% of the events have either been moved from the peak of the spectrum to the tail and vice versa \cite{Linssen:2012hp}.  Mass measurements of gauginos based on template fits of the energy distribution of the hadronically decaying final-state Standard Model bosons showed biases of approximately 1\% with such modifications \cite{Linssen:2012hp}. In slepton measurements based on the edges of the lepton energy distributions, shifts of up to 0.2\% in the measured mass were observed \cite{Linssen:2012hp}. With a more realistic study of the uncertainties in the reconstruction of the luminosity spectrum, this uncertainty was reduced by approximately one order of magnitude \cite{Battaglia:2013bha}. In contrast, the shift observed in the present analysis is only $6 \times 10^{-5}$ even with the conservative uncertainty estimate, and thus completely negligible. This is due to the construction of the observable used for the mass measurement, $M_C$, which is independent of the true collision energy, and thus very insensitive to variations in the luminosity spectrum. 

\section{Summary}

The potential of a 3 TeV linear $e^+e^-$ collider based on CLIC technology for the measurement of TeV-scale light-flavor right-handed squarks that decay into a quark and the lightest neutralino has been studied in full detector simulations including standard-model and beam-induced backgrounds in the framework of the CLIC CDR. Using a combination of missing energy and multivariate classifiers it was possible to achieve a high signal significance despite standard-model
background processes that exceed the signal production cross section by almost four orders of magnitude. The beam-induced background from $\gamma\gamma \rightarrow {\rm hadrons}$ processes is controlled by timing cuts in the reconstruction and by the use of the longitudinally invariant $k_t$ algorithm for jet finding. For  right-handed squarks with a mass of around 1125 GeV and a combined production cross section of 1.5 fb, a statistical precision of 6.5 GeV, corresponding to 0.58\%, was achieved for the mass measurement for combined up- and down-type squarks with an integrated luminosity of 2 ab$^{-1}$ at 3 TeV using a template fit with generator-level templates. Systematic changes in
the luminosity spectrum have a negligible effect on the measurement, demonstrating the robustness of the technique used for the mass measurement. Other systematics are expected to be comparable to or smaller than the statistical uncertainties. These results demonstrate the capability of CLIC for precision measurements of generic new TeV-scale states decaying into a hadronic jet and an invisible particle.

\begin{acknowledgements}
The work presented in this paper has been carried out in the framework of the CLIC detector and physics study, and we would like to thank in particular Stephane Poss of the CERN LCD group for the generation of the simulation samples used in this study as well as the CLICdp analysis working group for constant feedback on the analysis. The work was supported in part by the DFG cluster of excellence ``Origin and Structure of the Universe'' of Germany. 
\end{acknowledgements}

\bibliographystyle{spphys}
\bibliography{SquarksatCLIC}

\clearpage
\end{document}